\documentclass[a4paper,12pt,oneside]{article}
\usepackage[koi8-r]{inputenc}
\usepackage[english,russian]{babel}
\usepackage{graphicx}
\usepackage{amsmath}
\usepackage{amssymb}
\usepackage{euscript}
\begin{document}

\title{Results of magnetic field measurements in young stars \\DO Tau, DR Tau, DS Tau}

\author{A.\,V. Dodin$^1$, S.\,A. Lamzin$^1$, G.\,A. Chountonov$^2$}
\date{ \it \small
1) Sternberg Astronomical Institute of Moscow State University,
, Moscow, 119992, Russia\footnote{
{\it Send offprint requests to}: A. Dodin e-mail: dodin\_nv@mail.ru}\\
2) Special Astrophysical Observatory of the Russian AS, Nizhnij Arkhyz
369167, Russia\\
}

\maketitle

\bigskip

\section*{Abstract}
Results of measurements of the longitudinal magnetic field in a hot 
accretion spot in three classical T Tauri stars (CTTS) are presented.
The magnetic field in the formation region of the narrow component of 
the emission line He\,I 5876 {\AA} was found for each star in our 
sample at a level of more than 2$\sigma.$ In case of DS Tau we have 
found the field in the narrow components of Na\,I D lines, which was 
equal to $+0.8 \pm 0.3$ kG, i.e. it was equal to the field measured 
on the narrow component of He\,I 5876 \AA. Our results indicate that 
the magnetic field in the hot spots can be studied for CTTS down to 
13$^m$ that allow in the future to double a number of CTTS with measured 
field in the accretion zone.

\section*{ Introduction }

   Classical T Tauri stars (CTTS) are young ($t<10^{7}$ yr), low mass
$(M\leqslant 3\,M_\odot)$ stars at the stage of gravitational contraction
towards the main-sequence, activity of which is caused by magnetospheric
accretion of protoplanetary disc matter \cite{Bertout89}.
Inner regions of the accretion disc are truncated by the stellar magnetic 
field and  disc's matter  slides down toward the star along the field lines. 
Having reached the dense layers of the stellar atmosphere,  matter is 
decelerated in the accretion shock, behind the front of which the most 
part of its kinetic energy converts into the short-wavelength radiation.
One half of the radiation flux irradiates the star, producing the so-called hot spot on its surface.
Simulations, performed by Dodin \& Lamzin \cite{Dodin12b}, confirm the hypothesis, suggested by Batalha et al. \cite{Batalha96}, that the so-called narrow ($FWHM \sim 30$ km\,s$^{-1}$) components of emission lines in spectra of CTTS are formed in this region. Hence, if the narrow components of the emission lines are used for the magnetic field measurements, then the obtained field strength characterizes the field in the accretion zone on the stellar surface.

The second half of the short-wavelength radiation flux of the shock from the cooling zone escapes upward, heating and ionizing the pre-shock gas. Broad $(FWHM > 100$ km\,s$^{-1}$) components of the emission lines are formed in this region as well as in a more extended region at the truncation radius, where the disc interacts with the field, and a magnetospheric wind are launched.
The relation between the intensity of the narrow and broad components varies from line to line, changes over time, and for the same lines varies from star to star. Because the broad components are formed in the region at a significant distance from the stellar surface, the field strength, measured on these components, is, as a rule, in several times smaller than that, measured on the narrow ones (see, however, \cite{Dodin12a}).

The magnetic field of young stars determines an activity of CTTS and plays a crucial role in an evolution of angular momentum of these objects, therefore a question about strength and topology of the magnetic field is one of the fundamental questions of physics of young stars.
At the moment, the magnetic field has been found in about 40 CTTS.
These measurements are significantly differ in the methods, and therefore, the field strength, measured in different works, can relate to different spatial regions.
Maps of the magnetic field at the photospheric level have been reconstructed by analysis of polarized light for the following stars:
CV Cha and CR Cha \cite{Hussain09}, 
TW Hya          \cite{Donati11b},
V2129 Oph       \cite{Donati11a},
V2247 Oph       \cite{Donati10a},
AA Tau          \cite{Donati10b},
BP Tau          \cite{Donati08}.
 The field has been detected by means of Zeeman broadening in unpolarized light in 26 stars \cite{Johns-Krull07, Yang11}. The field on the narrow emission components, which are formed in the accretion zone, has been measured only in 10 stars:
RW Aur    \cite{Dodin12a},
TW Hya    \cite{Donati11b},
V2129 Oph \cite{Donati11a},
AA Tau    \cite{Donati10b},
BP Tau    \cite{Donati08, Symington05, Chuntonov07},
GM Aur, DF Tau, DN Tau, GG Tau  \cite{Symington05},
T Tau     \cite{Smirnov04}.

To measure a magnetic field,  spectra with a high quality are needed, therefore the observations carry out on large telescopes for bright stars with numerous strong lines.
However, the number of such stars is small, and practically, it is limited by the list, presented above.
If we do not aim to determine the magnetic field in the photosphere of the star, and want to determine the field only in the accretion spot at the stellar surface, by measuring the narrow components of the emission lines, then the success of the measurement is determined by the flux and shape of the profile of the lines, rather than by the brightness of the star.

Having considered a few dozen spectra of CTTS, obtained at the VLT and Keck observatories, we chose 8 relatively faint stars with the strong and narrow emission line of He\,I 5876, the magnetic field of which had never been measured before. In our paper the measurements for three of these eight stars are presented: 
DO Tau (spectral type M0, range of variability in $V=13^{m}.0-14^{m}.3$ \cite{Herbst94}), 
DR Tau (K5, $V=10^{m}.8-12^{m}.8$ \cite{Grankin07}), 
DS Tau (K5, $V=11^{m}.6-12^{m}.7$ \cite{Grankin07}).

			%%%%%%%%%%%%%%%%%%%%%%%%%%%%%%%%%%%%%%%%%%%%%

\section*{Observations and their reduction}

  The method we use to measure magnetic field is based on the fact that
so-called $\sigma$-components resulted in Zeeman splitting are polarized
circularly such as oppositely polarized components are located on different
sides of the central wavelength $\lambda_0$. If magnetic field in a
line formation region has a longitudinal component $B_z$ then the right-
and left-hand polarized components will be shifted relative to each other to \cite{Babcock58}:
$$
\Delta_B\simeq 2.3 \cdot 10^{-2} \,g\,
{\left( {\lambda_0 \over 5000} \right)}^2 \, B_z, \eqno(1)
$$
where $g$ is the Lande $g$-factor of the considered line,
$\Delta_B$ and $\lambda_0$ are in {\AA} and $B_z$ is in kG.
This relation allows to find the longitudinal magnetic field component $B_z,$
averaged over the line formation region, by measuring $\Delta_B$ from
two spectra observed in the right- and left-hand polarized light.

   Observations were carried out on 2012 October 26-27 with the Main 
Stellar Spectrograph\footnote{The description is aviable at: http://www.sao.ru/hq/lizm/mss/ru/tech.html (in russian)} \cite{Panchuk01} of 6-m telescope of Special Astrophysical Observatory equipped with a polarization $\lambda/4$-plate and a double slicer \cite{Chuntonov04}. The spectrograph slit width was $0.^{\prime\prime}5.$ The observations were carried out in the spectral range 5640-6480 {\AA} by using CCD array (E2V CCD42-90),  the size of which along the dispertion was 4600 pixels that corresponds to the inverse linear dispersion of  0.183 \AA/pixel.

Spectra were processed as follows \cite{Chuntonov07}. Night sky emission and detector bias as well as cosmic ray traces were removed in a standard way, using routines from the MIDAS software package. A spectrum of a thorium-argon lamp was used for a wavelength calibration. Each observed spectrum was transformed into the stellar rest frame by shifting it as a whole (by radial velocity) until a coincidence of photospheric lines in the observed and simulated spectra \cite{Dodin12b}.
The log of the observations is presented in Table \ref{tab-zhurnal}, which contains Julian Date JD of the middle of an observation, an exposure time and a signal-to-noise ratio S/N.

\begin{table}
\caption{Log of observations.}\label{tab-zhurnal}
\begin{center}
\begin{tabular}{c | c | c | c | c }
\hline\hline
 N & JD 245\,6220+ &Star  & $t_{exp},$ sec& S/N  \\
\hline
  1  & 7.181  & $\gamma$ Equ & 180  & 360 \\ % 59x60
  2  & 7.186  & $\gamma$ Equ & 180  & 360 \\ % 61x62
  3  & 7.299  & HD216228     & 100  & 180 \\ % 73x74
  4  & 7.324  & DS Tau       & 1200 & 62  \\ % 76x77
  5  & 7.354  & DS Tau       & 1200 & 64  \\ % 78x79
  6  & 7.383  & DS Tau       & 1200 & 60  \\ % 80x81
  7  & 7.420  & DR Tau       & 1200 & 123 \\ % 83x84
  8  & 7.449  & DR Tau       & 1200 & 119 \\ % 85x86
  9  & 7.479  & DR Tau       & 1200 & 113 \\ % 87x88
 10  & 7.510  & DO Tau       & 1200 & 83  \\ % 89x90
 11  & 7.540  & DO Tau       & 1200 & 78  \\ % 91x92
 12  & 7.628  & HD31398      & 100  & 110 \\ % 99x100
 13  & 7.638  & 53 Cam       & 300  & 370 \\ % 101x102
\hline
 14  & 8.212  & $\gamma$ Equ & 180  & 310 \\ % 22x23
 15  & 8.336  & DS Tau       & 1200 & 72  \\ % 33x34
 16  & 8.365  & DS Tau       & 1200 & 70  \\ % 35x36
 17  & 8.394  & DS Tau       & 1200 & 78  \\ % 37x38
 18  & 8.429  & DO Tau       & 1200 & 49  \\ % 41x42
 19  & 8.458  & DO Tau       & 1200 & 49  \\ % 43x44
 20  & 8.488  & DO Tau       & 1200 & 44  \\ % 45x46
 21  & 8.521  & DR Tau       & 1200 & 88  \\ % 49x50
 22  & 8.550  & DR Tau       & 1200 & 89  \\ % 51x52
 23  & 8.579  & DR Tau       & 1200 & 84  \\ % 53x54
 24  & 8.639  & 53 Cam       & 300  & 350 \\ % 58x59
\hline
\multicolumn{5}{p{9cm}}{\footnotesize N is a sequence number of an observation, which consists of two expositions, $t_{exp}$ is an exposure time of these expositions . S/N is a signal-to-noise ratio at a continuum level for each exposition.} \\
\end{tabular}
\end{center}
\end{table}

 To exclude systematic instrumental errors, our observations were organized as follows.
One measurement of the field needed two exposures, between which the superachromatic quarter-wave phase plate was rotated in such a way that the right- and left-hand polarized spectra changed places on the CCD array. Thus, we get four spectra of the star: right- and left-hand polarized at the first exposure $R_1$, $L_1$ and similar spectra of $R_2$, $L_2$ at the second exposure.

Further we calculated the difference between the positions of the lines in the spectra $R_1$ and $L_1$, which we denoted as $\Delta_1$, between $R_2$ and $L_2$, which we denoted as $\Delta_2$ and between $R_1+L_1$ and $R_2+L_2,$ which we denoted as $\Delta_3.$
Here $R+L$ denotes the sum of simultaneously obtained spectra. These differences were calculated by using the crosscorrelation method \cite{JP86} for the confidence level $\alpha=0.68,$ that corresponds to 1\,$\sigma$ error.
The shift due to Zeeman effect is equal to
$$
\Delta_B=\frac{\Delta_1+\Delta_2}{2},
$$
a mean systematic shift between $R$ and $L$:
$$
\Delta=\frac{\Delta_2-\Delta_1}{2},
$$
and a systematic shift between two exposures:
$$
\delta=\Delta_3.
$$
All shifts were calculated in pixels, and then, if necessary, were translated in \AA.
We used this method to measure $B_z$ on the lines He\,I 5876, Na\,I D and  [O\,I] 6300.

However, in the case of the field measurements on many lines, it is appropriate to measure the shifts all considered lines at once. In this case, the procedure of measurements $\Delta_B$ should be refined.
The shifts $\Delta_i$ $(i=1,2)$, calculated by the crosscorrelation method, include two shifts of different nature: $\Delta_B$ and $\Delta.$
On the one hand, the shift $\Delta_B$ is different for each spectral line, therefore to find $\Delta_i$ by calculating the maximum of the correlation function, we should choose the value of $B_z$ as an independent variable, because it is the same for all lines, while the shift of each line should be calculated by the formula~(1). On the other hand, the systematic shift $\Delta$ is assumed to be constant for the entire spectrum, therefore to find it, we should choose a shift of the spectrum as a whole as the independent variable.

To combine these contradictory requirements, we had found an average systematic shift $\overline{\Delta}$, assuming that it behaves like $\Delta_B$ and then we corrected the spectrum for this shift $\overline{\Delta}.$ Because $\Delta \ne \overline{\Delta},$ this correction does not eliminate the systematic error, but only reduces it. However, 3-4 iterations of the procedure allow to eliminate the systematic shift almost entirely. An uncertainty of the value $\Delta_B$, defined in this manner, equals to $0.5\,\sqrt{\sigma_{\Delta1}^2+\sigma_{\Delta2}^2}.$

Lande factors of photospheric lines were adopted from the VALD database \cite{Kupka99}.
In the case of blended photospheric lines $g$-factors were averaged with weights of a central depth:
$$
g_{ef}(\lambda) = \frac{\sum{g_if_i(\lambda)}}{\sum{f_i(\lambda)}},
$$
where $f_i(\lambda)$ is a gaussian profile of an $i$-th line in the blend. The width of the gaussian  was found from the observed spectrum, and the amplitude was equated to the line depth from the VALD database. 
We take into account all photospheric lines  with known Lande factors $g_i,$ which fall in our spectral range and have a central depth greater than 0.1. Lande factors for emission lines, considered in our paper: He\,I 5876 (excitation potential $\epsilon = 23.07$~eV),  [O\,I] 6300 ($\epsilon =1.97$~eV) and Na\,I 5890, 5896 ( $\epsilon =2.10$~eV) were assumed to be 1.1, 1.0, 1.33, correspondingly.

To test the discribed method, we have observed stars with a known magnetic field: 53 Cam and $\gamma$ Equ, as well as giants: HD\,216228 and HD\,31398, the field of which should be close to zero.
The results of the measurements $B_z$ are presented in the Table\,\ref{standart}.
It can be seen from the table that the values of $B_z$ are equal to zero within its uncertanties in case of 
HD\,216228 and HD\,31398 and, in case of $\gamma$ Equ, $B_z$ agree with the results, obtained by Kudryavtsev \& Romanyuk \cite{Kudryavtsev12}, who have found that for this star $B_z$ varies from $-0.85$ to $-1.25$ kG.
These results convince us of the correctness of the chosen method.
In case of 53 Cam, the result of second measurement coincides within the error with the expected value, calculated from the formulas from Hill's paper \cite {Hill98}, but $B_z$ of first measurement differs noticeably from the predicted value. Note that similar and even greater deviations from the predicted 
values were also observed in 53 Cam by other authors \cite{Kudryavtsev12}.

\begin{table}
\caption{Test measurements of the field}\label{standart}
\begin{center}
\begin{tabular}{c | c | c |c c| c }
\hline\hline
N  & JD 245\,6220+ & Star & $B_z$ & ~~~$\sigma_B$  & $B_{e}$ \\
\hline
1  & 7.181 & ~~$\gamma$ Equ~~ & $-1.00$ & ~~~$0.09$ &  $-1.1^{a}$ \\
2  & 7.186 & ~~$\gamma$ Equ~~ & $-0.94$ & ~~~$0.09$ &  $-1.1^{a}$ \\
3  & 7.299 & ~~HD 216228~~    & $-0.05$ & ~~~$0.04$ &  $ 0.0$ \\
12 & 7.628 & ~~HD 31398~~     & $-0.03$ & ~~~$0.04$ &  $ 0.0$ \\
13 & 7.638 & ~~53 Cam~~       & $-2.50$ & ~~~$0.09$ &  $-1.4^{b}$ \\
\hline
14 &  8.212 & ~~$\gamma$ Equ~~ & $-0.89$ & ~~~$0.09$ &  $-1.1^{a}$ \\
24 & 8.639 & ~~53 Cam~~       & $+1.92$ & ~~~$0.11$ &  $+2.1^{b}$ \\
\hline
\multicolumn{6}{p{11cm}}{\footnotesize
$B_z$ and $\sigma_B$ are the value of the measured field and its uncertainty, in kG.
$B_e$ -- an expected value of the field in kG:
(a) marks a mean value from the paper by Kudryavtsev \& Romanyuk \cite{Kudryavtsev12},
(b) marks a predicted value from the paper by Hill et al. \cite{Hill98}.
} \\
\end{tabular}
\end{center}
\end{table}

\section*{Results}

CTTS usually have a period of axial rotation about a week \cite{Artemenko12}, 
therefore it can be expected that the field could not change significantly during the time of the observation of every star within each night (around 2 hours). Hence, there are reasons to calculate an average $B_z$ for each night:
$$\overline{B_z} = \frac{\sum(B_{zi}/\sigma_{i}^{2})}{\sum(1/\sigma_{i}^{2})}.$$
An error $\sigma_{i}$ of an individual observation was estimated from the uncertanties of $\Delta_1,$ $\Delta_2$, as it was described in the previous section.

A scatter of the individual observations $B_{zi}$ during one night may be caused by the noise of the spectrum. In this case an uncertainty of the average can be estimated as:
$$
\sigma_{a}=\left[\sum(1/\sigma_{i}^{2})\right]^{-1/2}.
$$
However, we can not exclude that the scatter may be caused by some other random process, for which the standard deviation can be estimated as follows:
$$\sigma_{b} = t\left(n-1,1-0.5\cdot[1-P]\right)\times\sqrt{\frac{1}{(n-1)
\sum(1/\sigma_{i}^{2})}
\sum{\frac{(B_{zi}-\overline{B_z})^2}{\sigma_{i}^{2}}}},
$$
where $P=0.68$ is the confidence level corresponding to 1\,$\sigma,$
$n$ is the number of measurements being averaged, $t$ -- the inverse of Student's T cumulative distribution function. If there are both processes, and they are independent, then a total error of the average over night can be estimated as 
$\sigma_{\overline{B}} = \sqrt{\sigma_{a}^2+\sigma_{b}^2}.$
The values of $\overline{B_z}$ and $\sigma_{\overline{B}}$ are given in the Table.\,\ref{TTS}.

				%%%%%%%%%%%%%%%%%%%%%%%%%%%%%%%%%%%%%%%
\begin{table}
\caption{The results of the field measurements in CTTS}\label{TTS}
\begin{center}
\begin{tabular}{c | c | c c| c c | c c | c c}

Star                       &    N--N   &$\overline{B_z}$  &$\sigma_{\overline{B}}$ & $\overline{B_z}$ &$\sigma_{\overline{B}}$& $\overline{B_z}$&$\sigma_{\overline{B}}$& $\overline{B_z}$&$\sigma_{\overline{B}}$\\
                               &         &  \multicolumn{2}{|c}{He\,I 5876} & \multicolumn{2}{|c}{Na\,I D}& \multicolumn{2}{|c}{ 6300 [O\,I]} & \multicolumn{2}{|c}{The photosphere} \\
\hline
DO Tau    &  10--11  & $-0.79$ & ~~$0.27$ & $-0.13$ & ~~$0.30$ & $-0.01$ & ~~$0.31$ & $+0.23$ & ~~$0.29$\\
          &  18--20  & $-0.45$ & ~~$0.36$ & $+0.20$ & ~~$0.51$ & $+0.08$ & ~~$0.40$ & $-0.14$ & ~~$0.44$\\
\hline
DR Tau    &  7--9   & $ -0.94$ & ~~$0.32$ & $-0.29$ & ~~$0.46$ & $-0.16$ & ~~$0.36$ & $-0.11$ & ~~$0.45$\\
          & 21--23  & $-1.51$  & ~~$0.37$ & $-0.44$ & ~~$0.30$ & $-0.03$ & ~~$0.31$ & $-0.55$ & ~~$0.41$\\
\hline
DS Tau    &  4--6   & $+0.32$  & ~~$1.97$ & $-$     & ~~$-$    & $-0.28$ & ~~$0.63$ & $+0.05$ & ~~$0.70$\\
          & 15--17  & $+0.80$  & ~~$0.34$ & $-$     & ~~$-$    & $+0.56$ & ~~$0.70$ & $+0.02$ & ~~$0.64$\\
\hline
\multicolumn{10}{p{11cm}}{\footnotesize N--N is the sequence numbers from the Table\,\ref{tab-zhurnal}, $\overline{B_z}$ and $\sigma_{\overline{B}}$ are the value of longitudinal component of the field and its uncertainty, in kG.}
\end{tabular}
\end{center}
\end{table}

				%%%%%%%%%%%%%%%%%%%%%%%%%%%%%%%%%%%%%%%

\begin{figure}[p]
  \includegraphics[scale=0.7]{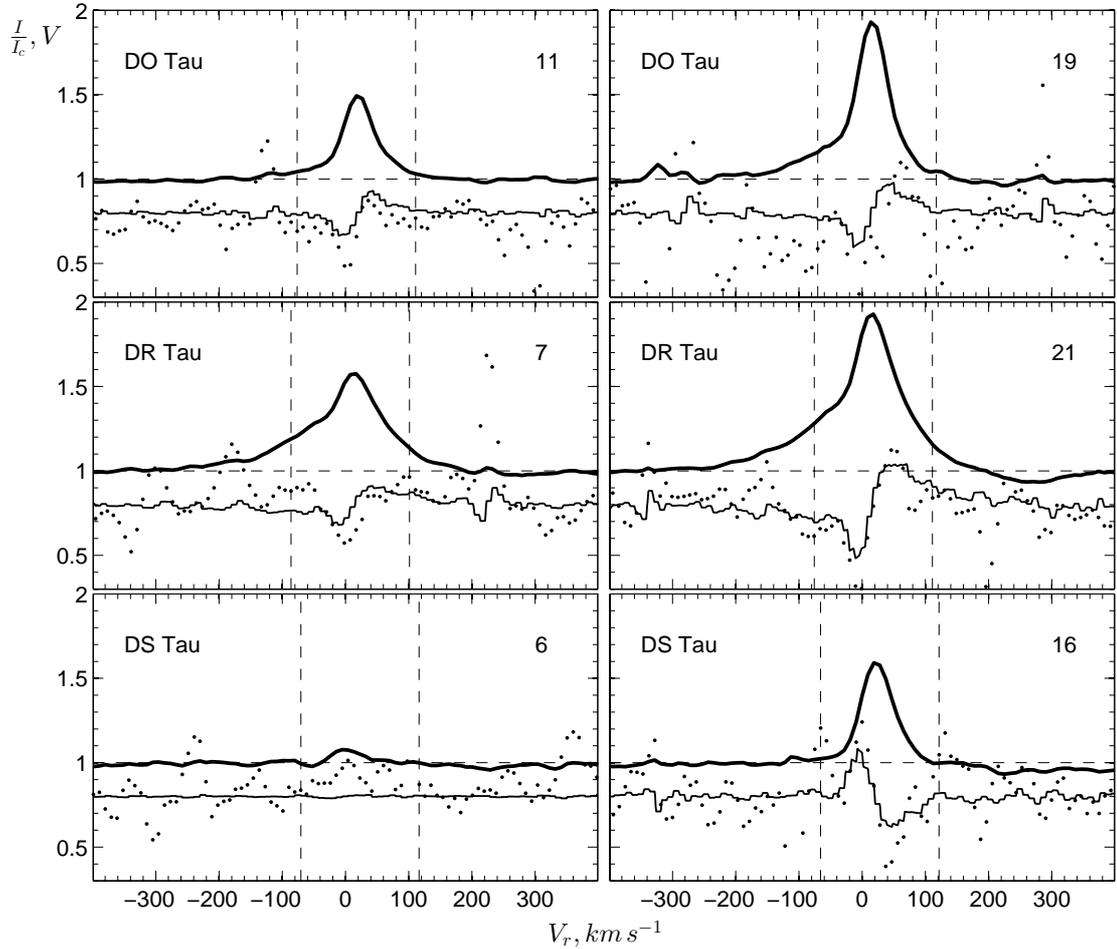}
  \caption{
Examples of $I-$ and $V-$profiles near the line He\,I 5876 for the stars DO Tau, DR Tau, DS Tau.
The left and right panels correspond to the observation on October 26 and 27, correspondingly.
The number in the upper right corner of each panel corresponds to the number in the Table\,\ref{tab-zhurnal}.
The thick curve indicates the observed $I-$profile. The dots indicate the observed $V-$profile.
The thin curve indicates the predicted $V-$profile for the field value from the Table\,\ref{TTS}. 
Both $V$-profiles were multiplied by 10 times and shifted to 0.8 for readability.
Dashed lines mark the continuum level and a spectral range, on which we measure the field.
An abscissa is the velocity in km\,s$^{-1}$ from the line center $\lambda_0=5875.6$ {\AA} in the stellar reference frame.}
\label{example}
\end{figure}

It follows from the Table \,\ref{TTS} that the fields have been detected in the formation region of the narrow component of the line He\,I 5876 in the stars DO Tau and DS Tau at a level $>2\sigma$ and in the star  DR Tau at a level $>3\sigma$. 
Examples of $I-$profiles for the given line for these stars are presented on the Fig.\,\ref{example}.
Vertical dashed lines mark the part of the profile (21 pixels), which was used in the measurement.
A variation of the equivalent width of the line from night to night can be caused either by the stellar rotation or by changes of physical conditions in the accretion spot. 
It manifested most clearly in the case of DS Tau: the line of He\,I 5876 in first night was approximately six times smaller than in second night.

				%%%%%%%%%%%%%%%%%%%%%

	  				%%%%%%%%%%%%%%%%%%%%%

			%%%%%%%%%%%%%%%%%%%%%

Dots on the figure correspond to an "observed"{} $V$-profile, which was calculated by using  the left- and right-polarized spectra, summarized pixel by pixel over the two exposures and corrected for $\Delta$ and $\delta:$
$$R = R_1(\lambda)+R_2(\lambda-\Delta-\delta),$$
$$L = L_1(\lambda-\Delta)+L_2(\lambda-\delta),$$
$$
V=2\frac{R-L}{R+L}.
$$
The shifts $\Delta$ and $\delta$ are about tenths of a pixel, therefore to calculate the values of $R_2$ and $L_2$ for fractional pixel values, we use a linear interpolation, which leads to a slight smoothing of the spectra. A thin line on the figure shows a "theoretical"\, $V$-profile, which was calculated as a relative difference of two $I$-profiles, the first of which was shifted by $\Delta_B/2$ to the left and the second was shifted by the same amount to the right.

\begin{figure}[p]
  \includegraphics[scale=0.7]{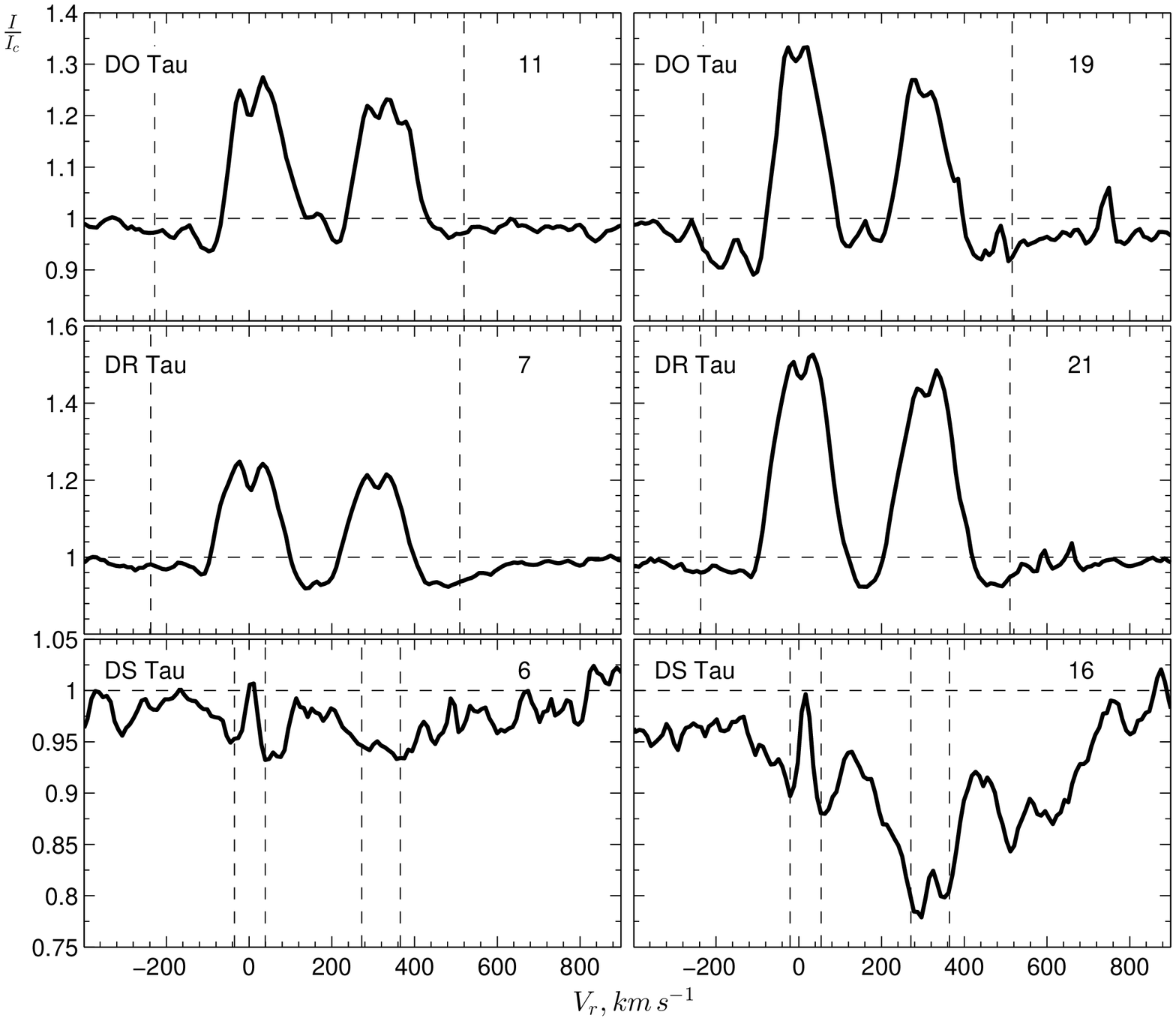}
  \caption{
Examples of $I-$profiles near the line Na\,I D for the stars DO Tau, DR Tau, DS Tau at the first night (the left column) and at the second night (the right corner).
The number in the upper right corner of each panel corresponds to the number in the Table\,\ref{tab-zhurnal}.
Dashed lines mark the continuum level and a spectral range, on which we measure the field.
An abscissa is the velocity in km\,s$^{-1}$ from the line center $\lambda_0=5889.95$ {\AA} in the stellar reference frame.
}
\label{NaI}
\end{figure}

In case of DO Tau and DR Tau the broad emission component (see Fig.\,\ref{NaI}) dominated in Na\,I D lines, the field on which was measured on both line simultaneously on the spectral range 5885.6 -- 5900.3 {\AA}.
Results of the measurements are collected in the the Table\,\ref{TTS} and they show that the value $\overline{B_z}$ in the broad emission components of Na\,I D is equal to zero with accuracy better than $\sim 1.5\sigma$ that is typical for the broad components of emission lines in CTTS (see the Introduction).

The broad emission component of Na\,I D lines is absent in DS Tau and the lines consist from broad photospheric absorption wings and the narrow emission component. The value of $B_z,$ measured on the narrow components and averaged over first night (numbers 4-6 from the Table.\,\ref{tab-zhurnal}), was equal to
$-0.21\pm 0.34$ kG. In other words, the field in a line formation region of Na\,I D as well as He\,I 5876 has not been detected in this night. However, similar measurements in the next night (the observations 15-17) gave the values $\overline{B_z}=+0.79 \pm 0.34$, which well coincided with $\overline{B_z},$ measured on He\,I 5876: $+0.80\pm0.34$ kG.

The field measured on the forbidden line of [O\,I] 6300 is always inside its uncertainty, because the line are formed at low gas densities in a large volume, where the magnetic field is close to zero.
The measurements of this line were carried out to check the absence of large systematic errors in the method of processing the spectrum.

The field, measured on photospheric lines, in all considered stars is equal to zero within its uncertainty.
This seems to be due to two effects.
First, we observed faint CTTS, and the signal-to-noise ratio for them was usually much lower than, for example, in case of the test stars, moreover, the radiation of the accretion spot leads to a significant shallowing the absorption lines in CTTS spectra (so-called veiling), that increases the error of $\overline{B_z}.$ Second, we measure only the longitudinal component of the field.
Hence, even if the field in the photosphere is the same as in the line formation region of He\,I 5876, then the longitudinal component of the field, averaged over the entire visible surface of the star,  is several times lower than in the line He\,I 5876, for which the field is averaged only over the accretion spot.
\cite{Johns-Krull99}.

			%%%%%%%%%%%%%%%%%%%%%%%%%%%%%%%%

\section*{Conclusion}

The possibility of determining the magnetic field in the accretion spot in a relatively faint CTTS by measuring the Zeeman splitting of the narrow component of strong emission lines in their spectra has been demonstated on examples of three stars. Following the tradition, we have measured the field on the line He\,I 5876, but lines of other elements, which as well as He\,I 5876 are formed in the accretion spot, can be also used in the future.

The measured values of $B_z$ are the values of the longitudinal component of the field, which is averaged over the accretion spot with weights of the line intensity in each point of the spot.
Dodin et al. \cite{Dodin13} have shown that there are lines in spectra of CTTS, the intensities of which vary in different ways depending on the accretion flux  $F_{ac} = \rho_0 V_0^3/2,$ where $\rho_0$ and $V_0$ are a pre-shock gas density and velocity, correspondingly. For instance, the intensity of Ca\,I lines is increased with increasing of $F_{ac},$ and vice versa Ca\,II lines get smaller.
It gives a possibility to use such lines for the measurements of the magnetic field in parts of the accretion spot with different values of the accretion flux $F_{ac}.$

\bigskip
We wish to thank The Large Telescopes Program Committee for providing time for the observations on 6-m telescope.  The observations on the 6-m telescope are held with a financial support of Ministry of Science and Education of the Russian Federation. The work was supported by the Program for Support of Leading Scientific Schools (NSh-5440.2012.2).

% - - - - - - - - - - - - - - - - - - - - - - - - - - - - - - - - - - - -

\end{document}